\documentclass[12pt,letterpaper]{article}
\usepackage{jheppub}
\usepackage{subfigure}

\title{Vacuum Structure and the Arrow of Time}

\author[a,b]{Raphael Bousso}
\affiliation[a]{Center for Theoretical Physics and Department of Physics,\\
 University of California, Berkeley, CA 94720, U.S.A.}
\affiliation[b]{Lawrence Berkeley National Laboratory, Berkeley, CA 94720,
  U.S.A.}

\abstract{We find ourselves in an extended era of entropy production.  Unlike most other observations, the arrow of time is usually regarded as a constraint on initial conditions.  I argue, however, that it primarily constrains the vacuum structure of the theory.  I exhibit simple scalar field potentials in which low-entropy initial conditions are not necessary, or not sufficient, for an arrow of time to arise.  I argue that the string theory landscape gives rise to an arrow of time independently of the initial entropy, assuming a plausible condition on the lifetime of its metastable vacua. The dynamical resolution of the arrow of time problem arises from the same structural properties of the string landscape that allow it to solve the cosmological constant problem without producing an empty universe, particularly its high dimensionality and the large difference in vacuum energy between neighboring vacua.}

\begin{document}
\maketitle

\section{Introduction}
\label{sec-intro}

The entropy $S$ of a closed macroscopic system either (a) increases or (b) is constant under unitary evolution. This is easily understood on statistical grounds: under reasonable assumptions, (a) is the most probable evolution of a system that has not yet reached its maximum entropy, $S_{\rm max}=\ln \dim \cal H$, where $\cal H$ is the Hilbert space of the system; while (b) is the most probable evolution of a system with $S=S_{\rm max}$.  

In theory, this statement is invariant under time reversal: Given a state at time $t$, the entropy will probably increase or stay constant, no matter whether we evolve in the past or future direction.  In practice, however, we observe that the entropy of matter systems increases only towards the future, never towards the past.  This asymmetry is called the (thermodynamic) arrow of time.  It can be explained by noting that the early universe was in a special state of relatively low entropy.  But why was this the case?  

This question is usually referred to the arrow-of-time ``problem'', but the arrow of time is an observation like any other: it can become a problem only in the context of a particular cosmological theory.  A theory is ruled out if its predictions conflict with the observed arrow of time.  And a theory is incomplete if it is compatible with but does not predict the observed arrow of time.

The purpose of this paper is twofold.  First, I will show that the conditions required for an arrow of time depend on the theoretical framework and cannot be equated with low initial entropy.   I will exhibit simple models that demonstrate that a low-entropy initial state is neither necessary nor sufficient for explaining the observed arrow of time.  

Secondly, I will argue that the landscape of string theory furnishes a powerful example in which the observed arrow of time can arise from arbitrarily high-entropy initial conditions.  This property rests on some of the same key features that allow the string landscape to solve the cosmological constant problem without fine-tuning.  It constitutes a second, independent success of the framework.


\paragraph{Outline and Summary}

In Sec.~\ref{sec-obs}, I characterize the observed arrow of time and distinguish it from other observations.  The arrow of time is defined as the entropy produced in our causal past since the time of big bang nucleosynthesis, $\Delta S\sim 10^{103}$.  

In Sec.~\ref{sec-one}, I consider cosmological models with a unique vacuum.  If the cosmological constant, $\Lambda$, is negative (Sec.~\ref{sec-neg}), one finds that the requirement of large maximum entropy leads to multiple challenges, in addition to the smallness of the initial entropy.  Most importantly, the magnitude of the cosmological constant must be small.  This requirement is shared by all other models considered later: because of the covariant entropy bound, the observed arrow-of-time alone requires the existence of a vacuum with unnaturally small cosmological constant.  In this sense, the cosmological constant problem is part of the arrow of time problem.  Of course, even if we set $-\Lambda\ll 1$, this class of models is ruled out by the observed positive sign of the cosmological constant~\cite{Rie98,Per98}.  In the remainder of the paper, I consider only models that contain at least one vacuum with positive vacuum energy.

In Sec.~\ref{sec-pos}, I still consider models with only one vacuum, but now the sign of its cosmological constant is taken to be positive.  Generic initial conditions lead to eternal inflation; and because of the absence of vacua with negative energy, all possible states recur infinitely many times, even in a single causal patch.  As was first pointed out by Dyson, Kleban, and Susskind~\cite{DysKle02}, this implies that the overwhelming majority of observers arise from the smallest possible fluctuations relative to the maximum entropy state, empty de~Sitter space.  Such observers are known as Boltzmann brains.  They observe no arrow of time other than the small arrow associated with their own existence.  This conclusion obtains independently of initial conditions, so these models conflict violently with the observed arrow of time.  They, too, are ruled out.

In Sec.~\ref{sec-two}, I consider small landscapes that contain both de~Sitter ($\Lambda>0$) and terminal ($\Lambda<0$) vacua, with initial conditions in one of the de~Sitter vacua.  This leads to eternal inflation.  In Sec.~\ref{sec-cp}, I review the associated measure problem and the causal patch measure.  In Sec.~\ref{sec-twovac}, a model with only two vacua is analyzed.  An arrow of time arises only if two conditions hold: the initial state must have low entropy, and the de~Sitter vacuum decay must proceed faster than the production of Boltzmann brains.  In Sec.~\ref{sec-domhis}, a fast method for reproducing these results is developed, based on the branching tree implementation of the causal patch measure.  

This method is applied to two models, each with four vacua, in Sec.~\ref{sec-four}.  The two models differ only through the ordering of the vacua; the values of the cosmological constant are the same.  Yet, in the first model, low-entropy initial conditions are necessary for an arrow of time; whereas in the second, an arrow of time is predicted even with high-entropy initial conditions.  

This constrast provides an instructive reference point for the analysis of the string landscape.  The key feature of the second model is that observers of any type---Boltzmann or ordinary---can be produced only after passing through a vacuum with large cosmological constant, and hence, very small entropy.  This feature is shared by the string landscape.

In Sec.~\ref{sec-big}, I claim that the string landscape predicts an arrow of time, independently of the initial entropy, if and only if all de~Sitter vacua decay faster than they produce Boltzmann brains.  Only four key properties of the string landscape, listed in Sec.~\ref{sec-properties}, are used in the proof of the claim, in Sec.~\ref{sec-proof}.  They are the high dimensionality of the landscape, the large but not double-exponentially large number of vacua, the lack of tuning, and the fact that neighboring vacua tend to have vastly different vacuum energy.  It is interesting that some of these properties also underly the solution of the cosmological constant problem.

\paragraph{Discussion}

The approach followed in this paper is based on semiclassical gravity.  The universe is assumed to have a beginning, and time is taken to end at spacelike singularities such as the big crunch in a vacuum with negative cosmological constant.  Whether these features survive in a full quantum gravity theory remains a major open question.  They are crucial here because they allow some of the models considered to evade ergodicity.  This is necessary for a large arrow of time to arise from arbitrarily high-entropy initial conditions.

All models considered except for the single vacuum in Sec.~\ref{sec-neg} exhibit the dynamics known as eternal inflation, and thus require a regulator or ``measure''.  Here the causal patch measure~\cite{Bou06,BouFre06a} is used, which is equivalent to the light-cone time cutoff~\cite{Bou09,BouYan09,BouFre10b}.  Results would be similar with closely related proposals such as the fat geodesic~\cite{BouFre08b,LarNom11}, scale factor~\cite{DesGut08a}, and CAH+ cutoffs~\cite{Vil11}.

In some of the literature on the arrow of time problem, it is taken for granted that the initial conditions of the universe should somehow be selected at random, from a Hilbert space that is imagined to be large enough to describe the presently visible universe.  To some~\cite{Wal05} (including myself, until recently), this seemed like a rather implausible assumption, since it was unclear how the observed arrow of time could arise dynamically from high entropy initial conditions.  In this paper, however, I demonstrate explicitly how this {\em can\/} happen.  Therefore, high entropy initial conditions are {\em not\/} automatically in conflict with observation.\footnote{This does not mean, of course, that high initial entropy should be taken for granted.  Since we have no clue as to the correct theory of initial conditions (if there is one), we have no right to exclude the possibility that the universe simply began small and/or with low entropy.  See, e.g., Refs.~\cite{Lin84b,Vil83b}.} Whether they are depends on the vacuum structure of the theory.

For example, the Hartle-Hawking no boundary proposal~\cite{HarHaw83} exponentially favors initial conditions corresponding to empty de~Sitter space with the smallest positive vacuum energy available in the theory.  In a theory with a single vacuum, this rules out the proposal.  But in certain potential landscapes such as the second model in Sec.~\ref{sec-four}, the Hartle-Hawking proposal is not excluded as a theory of initial conditions.  (In the landscape of string theory, the Hartle-Hawking proposal is not viable~\cite{BouZukTA}.\footnote{I would like to thank D.~Page for discussions of this point.})

\paragraph{Related work}

This paper builds on several results of earlier work, particularly on the seminal insights of Dyson, Kleban, and Susskind about recurrences in de Sitter vacua~\cite{DysKle02} (Sec.~\ref{sec-pos}), and on an analysis of the abundance of Boltzmann brains in the multiverse~\cite{BouFre08b,DesGut08b}.  Some of the assumptions and the proof in Sec.~\ref{sec-proof} differ slightly from those given in Ref.~\cite{BouFre08b}; in the context of the present paper, assumptions were desirable that distill out the key properties of the string landscape that lead to a dynamical explanation of the arrow of time.

The main points of the present paper are (i) the arrow-of-time problem is ill-defined outside a specific theoretical context; (ii) the vacuum structure of a theory is central to the question of whether and how an arrow of time arises; and (iii) the vacuum structure of the string landscape does lead to the prediction of an arrow of time, if a plausible condition on the lifetime of de~Sitter vacua is satisfied.

For a clear and concise introduction to the arrow of time problem, see Ref.~\cite{Wal05}.\footnote{The present paper provides counterexamples to two propositions discussed in Ref.~\cite{Wal05}: the claim that the arrow of time cannot arise dynamically but must come from initial conditions (it arises dynamically from high-entropy initial conditions in some models I exhibit, including the string landscape); and the claim that conditioning on the existence of observers cannot explain the arrow of time because it would be equivalent to assuming the arrow of time (we will encounter models in which such conditioning leads to the prediction of an empty universe except for a single Boltzmann brain, in conflict with the observed arrow of time).}  Interesting recent papers that relate to the arrow of time problem in eternal inflation or the string multiverse include Refs.~\cite{AlbSor04,CarChe04,Pag06,BouFre06b,Ban07,%
Alb09,CarTam10,AguCar11,Teg11,Nom11b}.

The present analysis is not directly related to the {\em Entropic Principle}~\cite{Bou06,BouHar07}.  Arguably, observation requires free energy.  Although this is not sufficient, matter entropy production turns out to be a successful proxy for observers in many settings~\cite{Bou06,BouHar07,CliFre07,BozAlb09,%
PhiAlb09,BouLei09,BouHar10}, and it has the advantage that allows for conditioning on observers without making specific anthropic assumptions such as the existence of galaxies, or stable nuclei, etc.  Employing the Entropic Principle for this purpose in no way prejudices the question of the arrow of time, since both Boltzmann brains and ordinary observers operate by increasing the entropy.  The question addressed here is why the entropy produced in the universe we see is much larger than necessary for our own operation: why is the arrow of time so large?

\section{The Observed Arrow of Time}
\label{sec-obs}

There are a number of choices in how one might characterize the observation that we call the ``arrow of time''.    It will be important to avoid conflating the arrow of time with other observations.  Imagine a fundamental theory that predicts a cosmology with the same amount of entropy production as in the universe we observe, but, say, by predicting the formation of black holes with half of the mass of a galaxy, in a small fraction of galaxies.  Or consider a theory that produced a lot of matter entropy, but through processes that are negligible in our universe, such as the decay of protons.  Each of these theories conflicts with observation, but not because they fail to predict an arrow of time.  

Similarly, the fact that we live at a time 13.7 Gyr after the big bang is an important observation---we would reject a theory that predicts that galaxy formation takes 100 billion years---but it is not among the observations that constitute the arrow of time.  To distinguish the arrow of time from other observations, I will define the observed arrow of time simply as the difference $\Delta S$ between the entropy $S(t_f)$ at the present time, $t_f$, and the entropy $S(t_i)$ at some earlier time, $t_i$:
\begin{equation}
\Delta S=S(t_f)-S(t_i).
\end{equation}

The entropy $S$ itself will be defined in the usual, coarse-grained way.  For example, the entropy density of radiation is of order $T^3$; the entropy of a dilute gas of particles is of order the number of particles, and the entropy of a black hole is its one quarter of its horizon area in Planck units.  

\begin{figure}[tbp]
\includegraphics[width=4in]{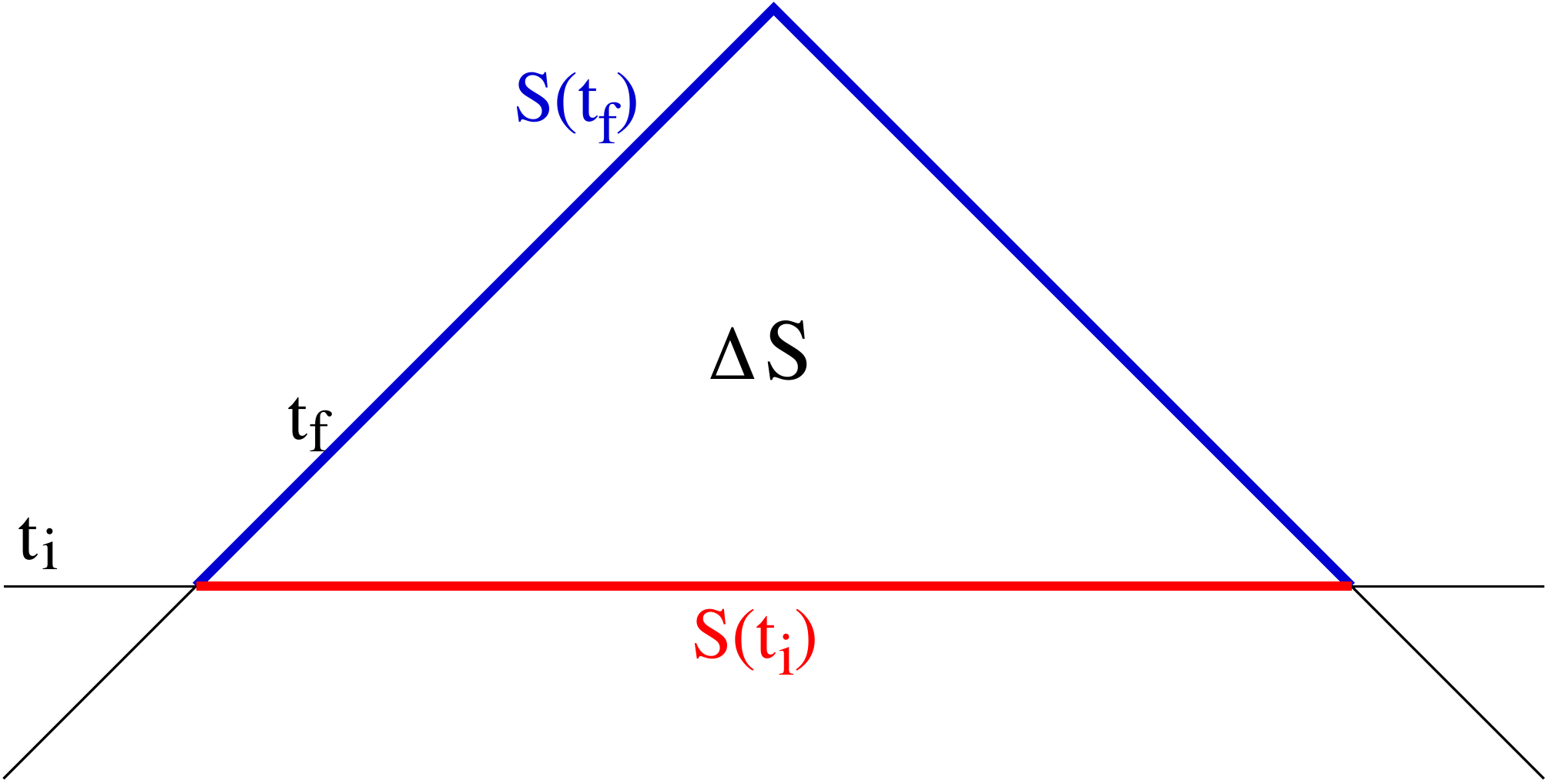}
 \caption{The observed arrow of time is defined as the entropy produced in our past light-cone (blue, $t_f$) since some early time that we can confidently extrapolate to, (red, $t_i$): $\Delta S=S(t_f)-S(t_i)$.  With $t_i$ equal to the era of big bang nucleosynthesis, one finds $\Delta S\sim 10^{103}$.}
\label{fig-arrowdef} 
\end{figure}
The next question is how to define $t_f$ and $t_i$.  I will define $t_f$ to be the observer's past light-cone (Fig.~\ref{fig-arrowdef}).  More precisely, $S(t_f)$ is the entropy on a null hypersurface defined as the intersection of the boundary of the observer's causal past with the future of $t_i$ (a spacelike hypersurface to be defined below). This choice has a number of properties that will be useful in the analysis below.\footnote{Two interesting alternatives are described in Ref.~\cite{EgaLin09}; they would yield qualitatively similar answers.  One option is to consider the entropy of a fixed comoving volume.  This relies on the special symmetries of FRW universes and involves an extrapolation to regions outside of our past light-cone.  Another possibility is to consider the entropy in the causal patch, which involves extrapolation to the future of the observer.}  The boundary of the past is covariantly defined, i.e., it does not depend on the choice of a time coordinate.  It is operational, in that it does not make reference to spacetime regions from which the observer could not have received any signals.  

Most importantly, the entropy  $S(t_f)$ can easily be bounded from above: the covariant entropy bound~\cite{CEB1,CEB2,RMP} guarantees that the entropy on any past light-cone cannot exceed its maximum area, in Planck units.   This leads to a convenient method for ruling out an arrow of time in large classes of models: we will encounter spacetime solutions that contain no past light-cones of sufficient size to allow for the amount of entropy production that we know to have taken place in our causal past.

We also have a choice about the ``initial'' time, $t_i$, whose entropy we compare to the present entropy.  I will take $S(t_i)$ to be the entropy on the corresponding timeslice (a timelike or null hypersurface), or more precisely, the portion of this slice that lies within the observer's causal past.  The choice of $t_i$ will be informed by the confidence we have about our understanding of ever more distant epochs.   None of the results in this paper are qualitatively sensitive to this choice.  

One (rather conservative) choice would be to let $t_i$ be 15 minutes ago, in the inertial frame of the Sun.  In this case, the arrow of time is quantified by the amount of entropy that was produced in your past light-cone roughly since you started reading this paper.  This is dominated by the photons emitted by the sun (during the first 7 minutes of that time-interval, since the sun is 8 light-minutes away and is thus outside our past light-cone at later times).  The entropy produced is of order the number of photons, $\Delta S\sim 2\times 10^{50}$.   

This choice of $t_i$ illustrates that the arrow of time is not associated only with cosmological time scales: a {\em decrease\/} in entropy of the same magnitude has probability of order $\exp(-10^{50})$.  The state of the universe 15 minutes ago was very special indeed compared to the current state, which in turn is very special compared to the state in 15 minutes.  Although Bekenstein-Hawking entropy dominates the entropy production on sufficiently large scales, this choice of $t_i$ also shows that the existence of a huge arrow of time does not require the presence of event horizons.

At the other extreme, one might choose as the initial time the Planck time, $t_i\sim 1$, defined as a hypersurface on which some curvature invariant becomes unity, in units of an appropriate power of the Planck length.  This choice has the disadvantage that the amount of entropy produced in our past light-cone since $t_i$ depends on physics in the early universe that we cannot be completely sure about.  For example, if today's CMB is the remnant of thermal radiation that was already present at a Planck-scale big bang, then it does not contribute to the entropy increase $\Delta S$.  But if the CMB were produced by reheating at the end of a era of slow-roll inflation, at some time $t_{\rm reheat}\gg 1$, then $\Delta S$ receives a large contribution from the CMB.

As a compromise between these extremes, I will take the initial time to be that of nucleosynthesis,
\begin{equation}
t_i=t_{\rm BBN}\sim 3~\mathrm{min}~.
\end{equation}
We have a rather firm understanding of the history of the universe since that time.  With this choice, the entropy produced, $\Delta S$, does {\em not\/} include the CMB, since the CMB was already present at the time $t_i$.  However, $\Delta S$ does include the Bekenstein-Hawking entropy of all black holes that formed after nucleosynthesis, which dominates in any case.  This includes both stellar-mass and supermassive black holes. The latter dominate the entropy production:\footnote{The numerical value is taken from Table 2, line 2, of Ref.~\cite{EgaLin09}.   Strictly, the value given there, $1.2^{+1.1}_{-0.7}\times 10^{103}$ refers to the supermassive black holes produced in the interior of the anticipated cosmological event horizon, which is slightly larger than our present causal past. The quantitative difference is not significant in the context of this paper.  Note that for reasons explained below, I am not including the cosmological horizon entropy, which would dwarf this value.}
\begin{equation}
\Delta S_{\rm SMBH}\sim 10^{103}~.
\end{equation}
This dwarfs the production of ordinary matter entropy, 
\begin{equation}
\Delta S_{\rm matter}\sim 10^{86}~,
\end{equation}
which is dominated by the production of the Cosmic Infrared Background (CIB): the down-scattering of optical stellar photons to the IR by galactic dust~\cite{BouHar07}.  

The results of this paper would be qualitatively unchanged, however, if black hole horizon entropy was excluded and only matter entropy was used to define $\Delta S$.  They would also be unchanged if one added to $\Delta S$ the horizon entropy of the cosmological horizon that will form around us in the future.  However, the correct choice is to include the Bekenstein-Hawking entropy of black holes, but to exclude the {\em cosmological\/} event horizon~\cite{GibHaw77a}, in the total entropy produced in our causal past since nucleosynthesis.  This distinction is based, once more, on what we are actually able to observe.  

Our past light-cone (more precisely, the boundary of our past) contains disconnected components that coincide, up to superexponentially small corrections, with the event horizons of black holes soon after those black holes form.  The matter that formed the black hole suffers exponential redshift with a characteristic timescale set by the radius of the black hole, which is much smaller than the age of the universe.  After a few times the light-crossing time of the black hole, a distant observer cannot receive signals from this matter, since such signals would take more energy to emit than the mass of the system and black hole combined~\cite{SusTho93}.   By causality, moreover, the observer is no longer able to travel to the matter and probe it directly.  Therefore such matter is entirely inaccessible and invisible to the observer, as is its entropy.  For the (generalized) second law of thermodynamics to hold, in this case, it is essential that we include the horizon entropy of the black hole~\cite{Bek72}.  

By contrast, the outermost component of the boundary of our past does not coincide with the cosmological event horizon, since we have not yet approached the asymptotic de Sitter regime, $t\gg t_\Lambda\sim 16$ Gyr.  If we lived in that era, it would be important to include the entropy of that horizon, $S\sim 3\times 10^{122}$.  But at the present era, signals that propagate to us from this outer component of the boundary of our past do not suffer exponential redshift; the matter is visible directly.  Another way of saying this is that the cosmological event horizon has not yet formed, the same way that we would look at a cloud of collapsing dust and say that it will form a black hole but has not yet done so.  In principle, any object that is just outside our past light-cone could still change direction and travel back to us (say, as a result of a collision), rather than crossing the cosmological horizon.

In summary, I take the observed arrow of time to be the entropy increase
\begin{equation}
\Delta S\sim 10^{103}~,
\label{eq-at}
\end{equation}
inferred from a combination of direct observation and theoretical modeling of the cosmological evolution since the time of nucleosynthesis.   

All I will take away from this section is this one number (and mainly, just the fact that it is very large).  Below, any model in which most observers see at least this amount of entropy production\footnote{The relevant $t_i$ may be defined as the latest time for which $\Delta S$ has a local maximum.  I am grateful to S.~Leichenauer for suggesting this definition.} (by some physical process that may vary from observer to observer) will be referred to as predicting an arrow of time.  And any model for which the majority of observers see much less entropy production will be considered in conflict with the observed arrow of time.  Either way, the models considered may be in conflict with other data, and this will occasionally be pointed out; but my focus will be on this one particular datum.

\section{The Arrow of Time in a Monovacuous Theory}
\label{sec-one}

In this section, I consider theories with a single, completely stable vacuum.  If the vacuum energy is negative, I find that low-entropy initial conditions are necessary but not sufficient for an arrow of time to arise, and I identify some of the additional necessary conditions.  If the vacuum energy is positive, however, then no arrow of time will be observed, independently of the initial entropy and any other conditions.

\subsection{Negative Cosmological Constant}
\label{sec-neg}

Consider a theory with a single vacuum with $\Lambda<0$, with a low-energy effective Lagrangian otherwise similar to our own.  Let us suppose that the theory sets not only the dynamical laws but also the initial conditions, and that it dictates that the universe comes into being as a flat, radiation dominated FRW universe, with Planckian density $\rho=1$ at the Planck time $t=1$ (I will use Planck units throughout).  I define
\begin{equation}
t_\Lambda\sim \sqrt{\frac{3}{|\Lambda|}}~,
\label{eq-tlb}
\end{equation}
the timescale associated with vacuum energy domination.

The Friedmann equations imply that the universe recollapses on a timescale of order
\begin{equation}
t_{\rm crunch}\sim t_\Lambda ~.
\end{equation}
As a result, the maximum area of any past light-cone in this universe is $t_\Lambda^2\sim |\Lambda|^{-1}$~\cite{BouFre10a}.  Thus, the entropy on an observer's past light-cone satisfies
\begin{equation}
S(t_f) \lesssim t_{\rm crunch}^2\sim |\Lambda|^{-1}~.
\end{equation}
Since the ininial entropy is nonnegative, the entropy increase $\Delta S$ satisfies the same inequality:
\begin{equation}
\Delta S \lesssim |\Lambda|^{-1}~.
\label{eq-dsneg}
\end{equation}

Thus we see that low-entropy initial conditions are not sufficient for an arrow of time to emerge.  It is also necessary that the cosmological constant be small in magnitude:
\begin{equation}
|\Lambda|\ll 1~.
\label{eq-neglambda}
\end{equation}
The same necessary condition will obtain in vacua with positive $\Lambda$ and in theories with multiple vacua.   With the observed value $\Delta S\sim 10^{103}$ from Sec.~\ref{sec-obs}, one obtains
\begin{equation}
|\Lambda|\lesssim 10^{-102}~.
\label{eq-102}
\end{equation}
Any model with larger $|\Lambda|$ necessarily conflicts with the observed $\Delta S$, simply because the maximum observable entropy is already smaller than $\Delta S$, independently of initial conditions.

This result is rather general---as general as the relation between $\Lambda$ and the maximum area of past light-cones~\cite{Ban00,Fis00b,Bou00a,Bou00b,BouDew02}.  It holds independently of the particle and field content of the theory and the details of initial conditions, including spatial curvature.  Of course, Eq.~(\ref{eq-102}) is far from the most stringent bound on $\Lambda$ one can obtain from observation~\cite{Wei89}.  The point is that it comes from the observation of the arrow of time {\em alone}, and it is already tight enough to pose a serious fine-tuning problem.  We may conclude that one cannot explain the observed arrow of time without explaining why the vacuum has so little energy. 

It should be stressed that low-entropy initial conditions and small vacuum energy, while necessary, are still not sufficient for an arrow of time.  We are accustomed to consider the arrow of time from the backward-looking perspective of an observer at late times.  But from a theory standpoint, it is more natural to take the forward-looking point of view.   Whatever the initial entropy is, the dynamical laws must allow for a large amount of entropy, $\Delta S\gg 1$, to be produced later on.  

This leads to other necessary conditions, such as the absence of large positive\footnote{The case of negative curvature is more subtle~\cite{Dys79}.} spatial curvature and the existence of massive particles.  With large positive curvature, $t_c\ll t_\Lambda$, the universe will recollapse on the timescale of curvature domination, $t_c$.  (By delaying $t_c$, slow-roll inflation can contribute to a successful prediction of an arrow of time.\footnote{Slow-roll inflation is occasionally criticized on the grounds that it ``worsens'' the arrow of time problem, since the universe at the beginning of inflation has even lower entropy than after reheating.  But this argument could be advanced against any theory that explains our observations by dynamical evolution from simple initial conditions (``simple'' in the sense of low complexity, as opposed to ``generic'', large-entropy initial conditions).  Examples include cold dark matter structure formation, galaxy formation, or big bang nucleosynthesis.  Indeed, by this standard, the theory that the universe was created an instant ago, with my memories included, should be given preference over all of standard cosmology with its enormous explanatory and predictive power.  The point is that we need to explain the arrow of time in any case, and there is no reason to believe that an explanation can be found for the very low entropy at nucleosynthesis but not for the very low entropy at the beginning of inflation.}) And without massive particles, the radiation era would last forever.  In that case, the evolution would be adiabatic, with $S(t_f)=S(t_i)$ and thus $\Delta S=0$.  A matter dominated era is crucial, as the expansion of the universe effectively creates new phase space (the empty space between massive particles) and thus room for entropy production.

By assuming an FRW universe that is initially radiation dominated, I imposed (relatively) low entropy initial conditions by hand.  This is fine for my present purpose, which was to demonstrate that even if low entropy initial conditions are granted, an arrow of time is not necessarily predicted.  

However, it is clear that in this model, low-entropy initial conditions are indeed necessary.  Let us assume that all other necessary conditions are satisfied.  In particular, assume that the cosmological constant is of equal magnitude as the observed value (but negative): $\Lambda\sim -10^{-123}$; and that the Lagrangian is otherwise the same as in our universe.  Now suppose that instead of a hot big bang, the initial conditions selected for a matterless universe dominated by large black holes, or a universe filled exclusively with thermal radiation of temperature 1000 K.  In this theory, observers would be extremely unlikely to arise.  And those that do (through rare downward fluctuations in the entropy) are overwhelmingly unlikely to observe a large arrow of time.  A more detailed argument would parallel the analysis of Boltzmann brains below.

Independently of the presence of an arrow of time, the class of models considered in this subsection are in conflict with observation~\cite{Per98,Rie98}, by virtue of having negative $\Lambda$.  Therefore, I will not consider such models further.  In the remainder of this paper, the fact that the observed sign of $\Lambda$ is positive (though not necessarily its precise value) will be used as a key constraint on the theoretical models considered.  As we shall see, the positive sign of $\Lambda$ considerably alters and sharpens the challenge of explaining the arrow of time.

\subsection{Positive Cosmological Constant}
\label{sec-pos}

I will now consider the same type of theory as before, except that the cosmological constant is positive: $\Lambda>0$.  I still assume that initial conditions select for a hot, spatially flat, radiation dominated big bang.
At late times, $t\gg t_\Lambda\equiv (\Lambda/3)^{-1/2}$, the universe asymptotes to de~Sitter space. 

In an asymptotically de~Sitter big bang cosmology, the maximum area of any past light-cone will not exceed the cosmological horizon area of empty de~Sitter space, $12\pi/\Lambda$~\cite{Ban00,Bou00a}.  The covariant entropy bound thus implies $S(t_f)\lesssim t_\Lambda^2$.  Thus, one finds again that an arrow of time, $\Delta S\gg 1$, can exist only if $\Lambda\ll 1$.  The magnitude of the cosmological constant must be small in order for the maximum observable entropy to be large, which in turn is necessary for a large entropy difference $\Delta S$.

However, {\em the probability of the observed arrow of time is double-exponentially small in this model}, even if all necessary conditions are satisfied.  This result is due to Dyson, Kleban, and Susskind (DKS)~\cite{DysKle02}.  It implies that our vacuum is only metastable.  It will eventually decay, though this may happen at a time exponentially greater than the present age of the universe.  It is remarkable that this powerful conclusion can be drawn from such sparse assumptions.

To understand this result, let us restrict attention to a causal patch: a single de~Sitter horizon volume, or more precisely, the interior of the cosmological event horizon.  This region has finite maximum entropy, given by the entropy of empty de~Sitter space,
\begin{equation}
S_{\rm dS}=\frac{3\pi}{\Lambda}~.
\end{equation}
The total entropy is lower when matter is present, as the additional matter entropy is always overcompensated by a decrease in the cosmological horizon area~\cite{Bou00a,Bou00b}:
\begin{equation}
S_{\rm bulk}+S_{\rm CH}<S_{\rm dS}~.
\end{equation}
Thus, de Sitter space can be regarded as a quantum-mechanical system with finite entropy.   The equilibrium state is empty de~Sitter space; states with matter or black holes have lower entropy.

Assuming ergodicity and unitarity of the underlying quantum theory, these properties imply that a configuration with coarse-grained entropy $S_{\rm bulk}+S_{\rm CH}$ occurs, on average, at a rate per unit time of order
\begin{equation}
\Gamma\sim t_\Lambda^{-1} ~\frac{\exp(S_{\rm bulk}+S_{\rm CH})}{\exp(S_{\rm dS})}~.
\label{eq-gammads}
\end{equation}
Here $S_{\rm bulk}$ includes the entropy of all matter systems and black holes within the cosmological horizon, and $S_{\rm CH}$ is one quarter of the cosmological horizon area in the presence of this bulk configuration.

If the cosmological horizon area is nearly that of empty de~Sitter space, $1-(S_{\rm CH}/S_{\rm dS})\ll 1$, then the bulk configuration can be assigned an approximate notion of energy, $E$. (Intuitively, $E$ is the integral over the stress tensor, plus the mass of small black holes; see Ref.~\cite{Bou00b} for details.)  In the same limit, an inertial particle detector will see a nonzero temperature, whose minimum is attained in empty de~Sitter space:
\begin{equation}
T_{\rm dS}=\frac{1}{2\pi t_\Lambda}
\end{equation}
The relation between energy and cosmological horizon area implies~\cite{Bou00b} that Eq.~(\ref{eq-gammads}) reduces to the usual Boltzmann suppression factor,
\begin{equation}
\Gamma\sim t_\Lambda^{-1}~\exp(S_{\rm bulk}-E/T_{\rm dS})
\end{equation}
in this regime. 

The DKS argument relies only on the fact that the rate in Eq.~(\ref{eq-gammads}) is nonzero.  For the sake of argument, let us suppose not only that there is a beginning of time (a big bang), but also that initial conditions and dynamical laws lead to a standard cosmology compatible with our observations. ``Ordinary observers'' like ourselves, who live in this initial period, will observe a large arrow of time.  However, at times much greater than $t_\Lambda$, the universe becomes empty de~Sitter space and the thermal description becomes valid.  The finite period of standard cosmology is followed by an infinite era of fluctuations and recurrences.

For every ordinary observer produced 13.7 billion years after the big bang, there will be infinitely many observers produced later by thermal fluctuations.  Thus, the transients following the big bang are irrelevant for the predictions of the theory, and the relative probability of observations of type 1 and type 2 are determined entirely by the relative frequency of these events.  From Eq.~(\ref{eq-gammads}), one finds 
\begin{equation} 
\frac{p_1}{p_2}=\frac{\exp[S_{\rm bulk}(1)+S_{\rm CH}(1)]}{\exp[S_{\rm bulk}(2)+S_{\rm CH}(2)]} 
\end{equation}
Now, let state 1 be the coarse-grained state we observe, with galaxies and dark matter and dark energy.  Let state 2 be some state with larger entropy but the same energy (more precisely, the same cosmological horizon area).  For example, state 2 might be a state with a CMB temperature of 4 K instead of 2.7 K, and a very slightly smaller number of protons.  State 2 has larger entropy due to the hotter CMB; the correction due to the the missing protons is negligible.  Since $S_{\rm CH}$ is the same for both states, their relative probability will be given by $p_1/p_2\sim \exp[S_{\rm bulk}(1)-S_{\rm bulk}(2)]\sim 10^{88}$~\cite{BouHar10,EgaLin09}.  Thus, our observations have double-exponentially small probability, and the theory is ruled out at an extremely high level of confidence.

State 2 would not arise by semiclassical evolution from simple initial conditions.\footnote{This is the reason why the arrow of time, $\Delta S$, is small, even though $S(t_f)$ is larger than in our universe.}  But it dominates over state 1 in the thermal ensemble, and this counts in the long run.  Moreover, state 2 is itself highly unlikely compared to state 3, in which the CMB temperature is 6 K.  Still vastly more likely would be state 4, which is empty de~Sitter space except for a single solar system containing the observer. Ultimately, the most probable observations result from the smallest fluctuations above empty de~Sitter space that barely suffice to produce observers.  The most probable observers do not see a large arrow of time.

One can only speculate~\cite{Ban07} about the minimum requirements for such ``Boltzmann brains''.  But the point is that they come at the end of a long list of coarse-grained states, all of which contain observers that see a universe totally different from ours, and all of which are overwhelmingly more probable than a state compatible with a long semi-classical history, such as the state we observe.

Therefore, theories with a single vacuum are ruled out by combining two simple observations: the arrow of time and of the positive sign of the cosmological constant.  More generally, the same argument rules out any theory that contains only de~Sitter vacua but no terminal vacua (i.e., no vacua with $\Lambda\leq 0$).\footnote{If thermally produced black holes in de~Sitter space act as terminal vacua, then there are no such theories~\cite{BroPC}.}

\section{The Arrow of Time in a Small Landscape}
\label{sec-two}

\subsection{Eternal Inflation and the Causal Patch Measure}
\label{sec-cp}

Eternal inflation arises in any theory with at least one long-lived metastable de~Sitter vacuum and with initial conditions that assign nonzero probability to a state that can evolve to such a vacuum.  The de~Sitter vacuum may decay,\footnote{The measure problem also arises in a theory that contains at least one {\em stable\/} de~Sitter vacuum. In its most general form, the causal patch cutoff can be applied to this problem, by terminating the evolution at the first recurrence~\cite{Bou06}.  This measure was implicit in the previous section and in Ref.~\cite{DysKle02}.}  but globally its volume continues to grow indefinitely~\cite{TASI07}.  The expected number $\langle N_I \rangle$ of occurences of any type $I$ of observation or experimental outcome, no matter how unlikely, is infinite unless it is completely forbidden.  This means that relative probabilities cannot be defined by 
\begin{equation}
\frac{p_I}{p_J}=\frac{\langle N_I \rangle}{ \langle N_J \rangle}~. 
\label{eq-relprob}
\end{equation}

The measure problem is the question of how to regulate this divergence and obtain well-defined probabilities.   Current proposals have focussed on geometric cutoffs, i.e., simple prescriptions for selecting a finite subset of the eternally inflating spacetime.  Then $\langle N_I \rangle $ and $\langle N_J \rangle$ can be computed in a finite region and relative probabilities can be defined by Eq.~(\ref{eq-relprob}).

Two leading proposals are the causal patch cutoff~\cite{Bou06} and the fat geodesic cutoff~\cite{BouFre08b,LarNom11}.   For definiteness, the causal patch cutoff will be used here.  However, because of the double-exponential nature of the relative probability for an arrow of time vs. no arrow of time (i.e., the ratio of ordinary observers to Boltzmann brains), both cutoffs agree~\cite{BouFre08b} on all issues explored in this paper.

The causal patch cutoff is defined as a single causally connected region of spacetime.  More precisely, it is the causal past of an inextendible geodesic in the eternally inflating spacetime.  This prescription is quite general and does not assume that de~Sitter vacua are extremely long-lived, or that vacua must be sharply distinguishable.   In all models considered in this paper, however, both of these conditions hold.  In this case, the problem of evaluating $\langle N_I \rangle $ in the causal patch can be broken up into two separate calculations: (1) the expected number of times $e_i$ that the geodesic will enter vacuum $i$; and (2) the expected number $N_{Ii}$ of events of type $I$ that will happen in vacuum $i$ after the geodesic enters it:
\begin{equation}
\langle N_I \rangle = \sum_i N_{Ii} e_i
\label{eq-break}
\end{equation}

The expected number of times the geodesic enters vacuum $i$, $e_i$, can be computed in terms of branching ratios.  If the geodesic is some vacuum $j$,  what matters is which vacuum $k$ will nucleate next (as a bubble within the causal patch).  The total decay rate of vacuum $j$, $\Gamma_j\equiv \sum_k \Gamma_{kj}$, is irrelevant.  The branching ratio from vacuum $j$ to $k$ is defined as
\begin{equation}
\beta_{kj}\equiv \frac{\Gamma_{kj}}{\Gamma_j}~.
\label{eq-branch}
\end{equation}
The expected number of times that a geodesic starting in some initial vacuum $*$ will enter vacuum $i$ is given by a sum over all possible decay paths from $*$ to $i$, of the product of branching ratios along the decay chain:
\begin{equation} 
e_i=\sum_{\mathrm{*\to\ldots\to i}} \beta_{i\ldots}\ldots\beta_{\ldots *} 
\end{equation}
Graphically, the set of all decay paths can be represented by a branching tree with an initial node at $*$~\cite{Bou06}.  

In de~Sitter vacua which are long-lived, the boundary of the causal patch agrees (up to double-exponentially small corrections) with the cosmological event horizon that would obtain if the vacuum were completely stable.  This will be the case for all de~Sitter vacua considered in this paper.  Then the causal patch cutoff simplifies to the statement that for each de~Sitter vacuum that is encountered by the geodesic, $N_{Ii}$ is given by the expected number of events of type $I$ that are counted are those taking place within the de~Sitter event horizon surrounding the geodesic.

\subsection{A Landscape with Two Vacua}
\label{sec-twovac}

Consider a theory with two vacua, $A$ and $T$.  If both are de~Sitter, then the analysis is the same as for a theory with a single de~Sitter vacuum (Sec.~\ref{sec-pos}).  There will be no arrow of time, independently of initial conditions.  If both have $\Lambda<0$, or one of them has $\Lambda<0$ and initial conditions select this vacuum, then the analysis is the same as for a theory with a single $\Lambda<0$ vacuum (Sec.~\ref{sec-neg}), and there will be an arrow of time subject to the necessary conditions of Eq.~(\ref{eq-neglambda}).   In any case, I am taking the observed positive sign of the cosmological constant as a constraint and consider only theories that have vacua with $\Lambda>0$.   Hence, I will consider the case where $\Lambda_A>0$, $\Lambda_T<0$, and initial conditions are in the basin of attraction of the de~Sitter vacuum.
\begin{figure}[tbp]
\centering
\subfigure[Low initial entropy, fast decay]{
   \includegraphics[width=2.5in]{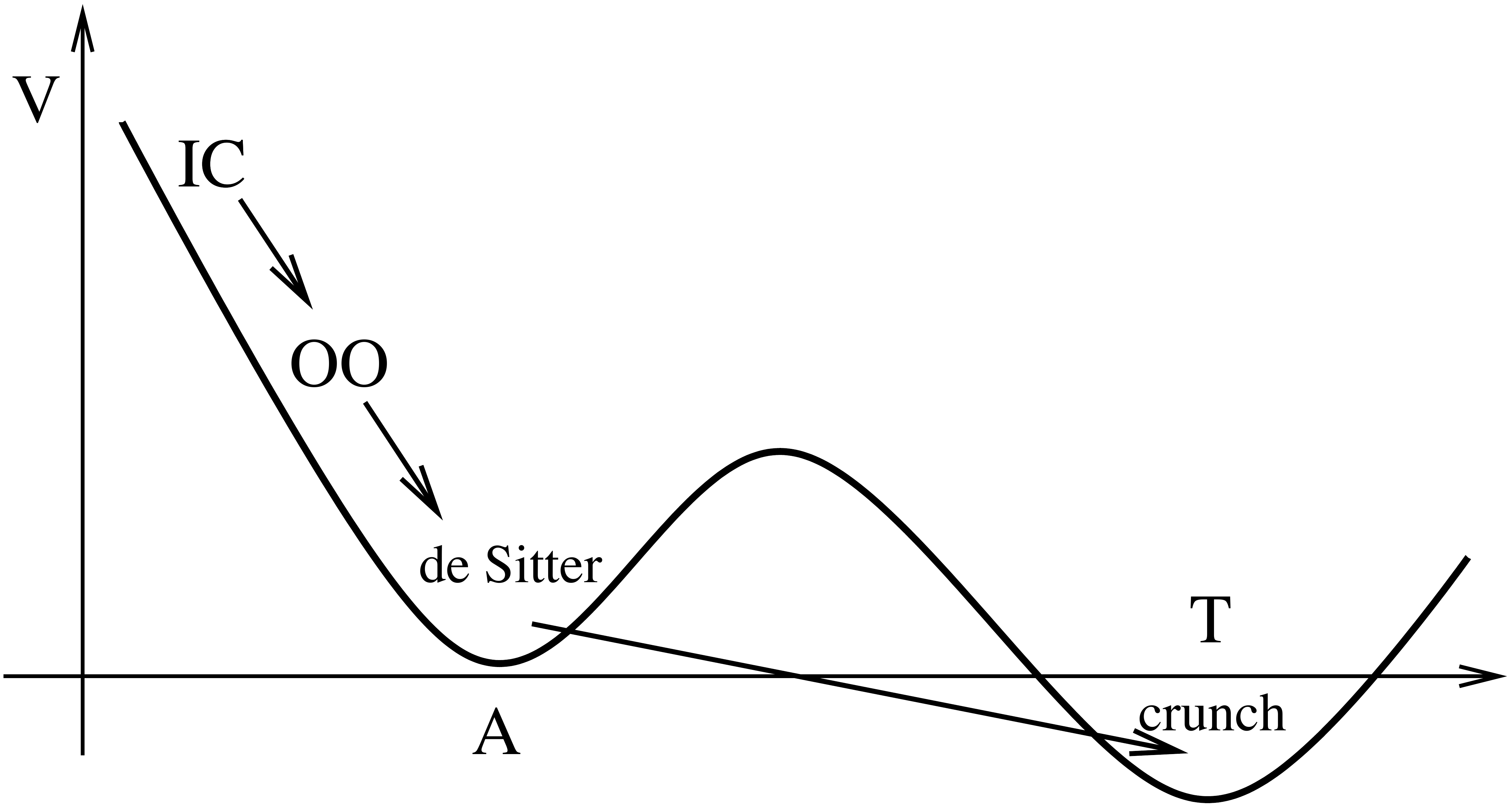}
   }
   \hspace{0in}
\subfigure[High initial entropy, fast decay]{
  \includegraphics[width=2.5 in]{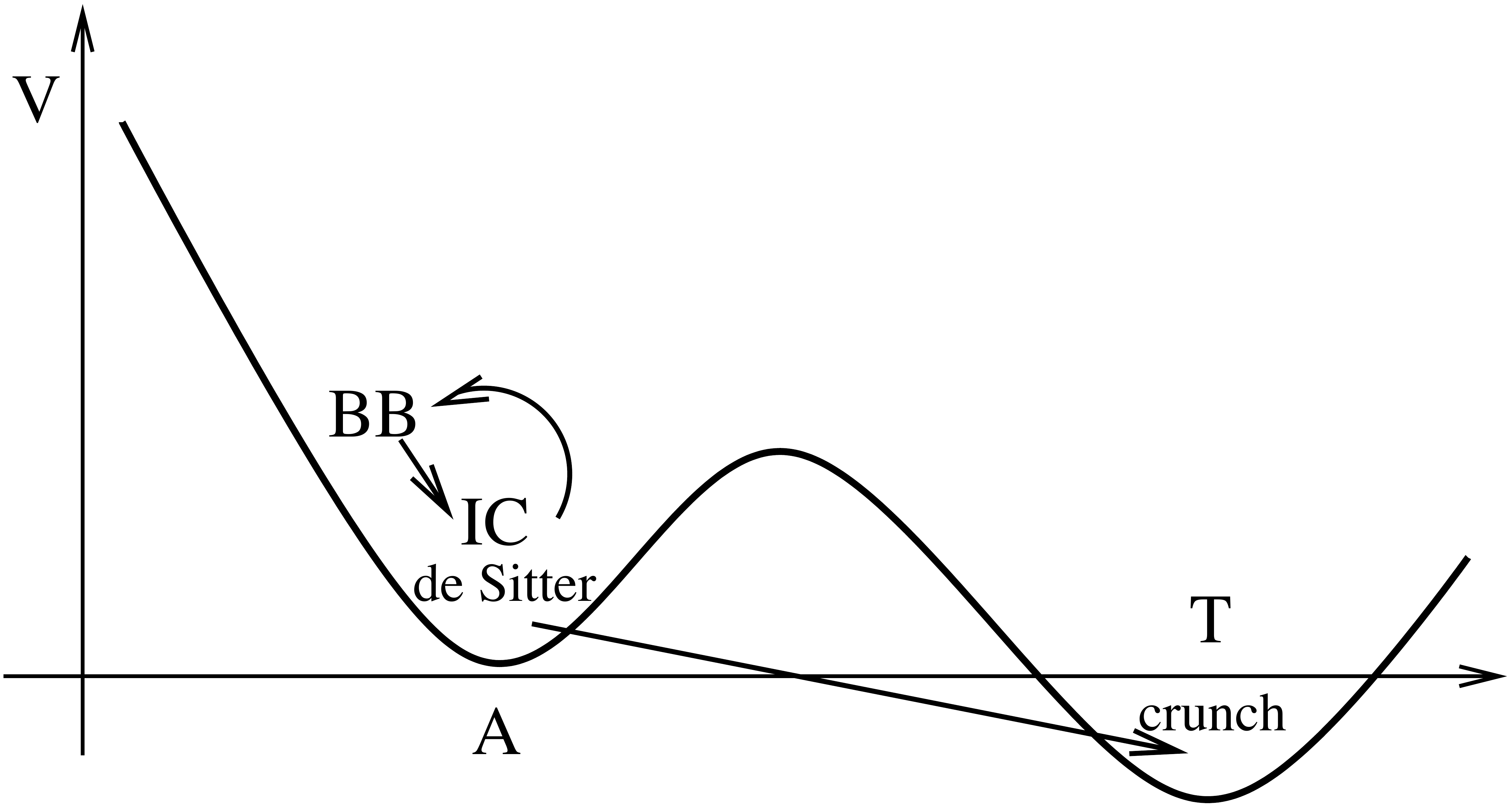}
   }
  \hspace{0in}
\subfigure[Low initial entropy, slow decay]{
  \includegraphics[width=2.5 in]{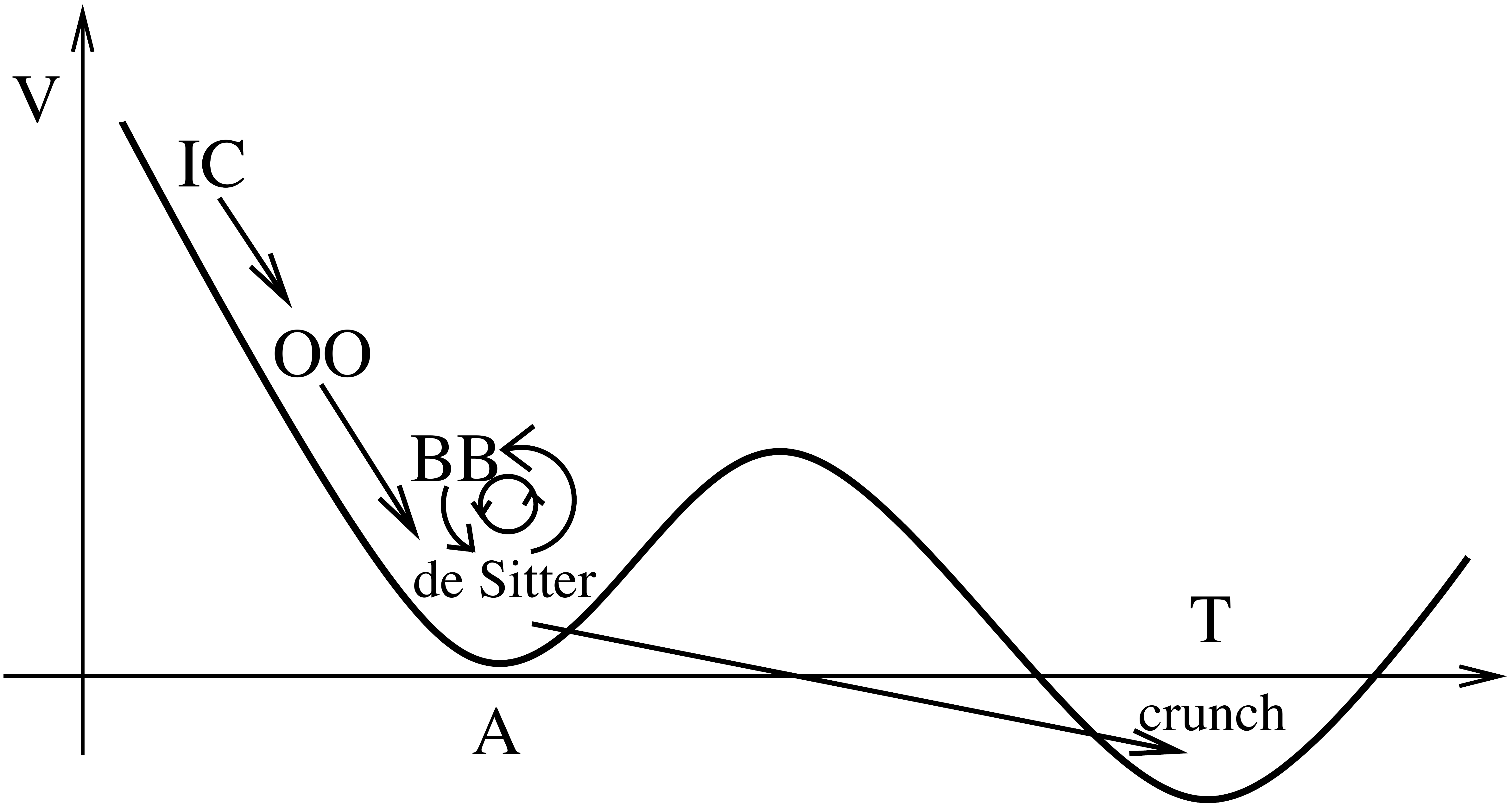}
   }
\caption{A landscape with one metastable de~Sitter vacuum $A$ and a terminal vacuum $T$.  The vertical axis indicates both the effective potential and, schematically, the free energy: ordinary observers, who see a long arrow of time, are drawn higher up than Boltzmann brains.  The dominant histories are indicated by chains of arrows.  (a) With  $\Gamma_A>\Gamma_{{\rm BB},A}$ and initial conditions (IC) similar to our own universe, most observers are ordinary observers (OO) who perceive an arrow of time. (b) If initial conditions are high-entropy, then the most likely observers are Boltzmann brains.  (c) If vacuum $A$ decays too slowly, ($\Gamma_A<\Gamma_{{\rm BB},A}$), then many more Boltzmann brains than ordinary observers are produced.}
\label{fig-AT} 
\end{figure}

The condition 
\begin{equation}
\Lambda_A\ll 1
\end{equation}
is necessary for an arrow of time, $\Delta S\gg 1$, since the maximum entropy in a de~Sitter vacuum is given by
\begin{equation}
S_A=\frac{3\pi}{\Lambda}
\end{equation}
But even if this condition is satisfied, the overwhelming fraction of observers will be Boltzmann brains (i.e., will not observe an arrow of time), unless the production rate of Boltzmann brains is smaller than the decay rate of vacuum $A$~\cite{BouFre06b}:
\begin{equation}
\Gamma_{{\rm BB},A}<\Gamma_A~.
\label{eq-fastdecay}
\end{equation}
If the above condition is not satisfied, then the probability of our observations (a universe rich with matter and radiation, and consistent with evolution from a low-entropy initial state) will be virtually nil, and the theory will be ruled out.

To understand this result, it will be necessary to review the arithmetic of non-finetuned exponentials and double-exponentials.  A number of the form $\exp(x)$ (or $\exp(-x)$), with $x\gg 1$, is called exponentially large (or small).  Numbers of the form $\exp[\exp(x)]$ (or $\exp[-\exp(x)]$), $x\gg 1$, are called double-exponentially large (or small).  Two exponentially large numbers $e_1>e_2$ satisfy $e_1\pm e_2\approx e_1$, unless their exponents are finetuned, i.e., as long as $|\ln e_1 - \ln e_2|\gg 1$.  Therefore, if $E_1$ and $E_2$ are both double-exponentially large and $E_1>E_2$, then in the absence of finetuning  (i.e., if $|\ln\ln E_1-\ln\ln E_2|\gg 1$), we have 
\begin{equation}
E_1 E_2\approx \frac{E_1}{E_2} \approx E_1~.
\label{eq-de}
\end{equation}
For a double-exponentially small number $E_3$, $E_4=1/E_3$ will be double-exponentially large and the above rule can be applied to it.  I will occasionally use the notation $x\lll y$ to emphasize that $y/x$ is double-exponentially large (note that only one of the two numbers needs to be double-exponentially large or small for this property to hold).  However, I will not necessarily use this notation when the fact that $y/x$ is double-exponentially large is obvious.  For example, if $y$ and $x$ are both double-exponentially large and mutually unrelated by fine-tuning, and $y>x$, then it follows trivially that $y\ggg x$.

The expected number of ordinary observers, $N_{\rm OO}$, will depend on the theory and on initial conditions.  In any case, however, it will be bounded by the number of particles that can fit within the causal patch, which is of order $\Lambda_A^{-1}$~\cite{BouFre10a}.  I will assume that $\Lambda_A$ is at most exponentially small (such as the observed value), but not double-exponentially small. Therefore, $N_{\rm OO}$ will not be double-exponentially large.  I will assume that $N_{\rm OO}$ is not double-exponentially small, though the analysis below could easily be augmented to include this case.  However, $N_{\rm OO}$ could be exactly zero.  For example, initial conditions or the dynamics of the theory may not give rise to ordinary observers even if Boltzmann brains are possible.

The expected number of Boltzmann brains in the causal patch is given by the lifetime of the de~Sitter vacuum times the rate at which Boltzmann brains are produced:  $N_{\rm BB}=\Gamma_{\rm BB}/\Gamma_A$.   I assume that $\Gamma_{\rm BB}$ is double-exponentially small but nonzero; otherwise there would be no observers of any kind, and the theory would be trivially ruled out.  $\Gamma_A$ is nonzero, since the de~Sitter vacuum is not completely stable by assumption.  $\Gamma_A$  may or may not be double-exponentially small. 

By the laws of double-exponential arithmetic,
\begin{equation}
\frac{p_{\rm OO}}{p_{\rm BB}}=\frac{N_{\rm OO}}{N_{\rm BB}}=\frac{N_{\rm OO} \Gamma_A}{\Gamma_{\rm BB}}\approx 
\left\{\begin{array}{rcl}
0 & \mathrm{~~if~~} & N_{\rm OO}=0 \\
\Gamma_A & \mathrm{~~if~~} & \Gamma_{\rm BB} > \Gamma_A~,~N_{\rm OO}>0 \\
\Gamma_{\rm BB}^{-1} & \mathrm{~~if~~} & \Gamma_{\rm BB} < \Gamma_A~,~N_{\rm OO}>0  
\end{array}\right.~.
\end{equation}
In the first and second case, the theory is ruled out because it conflicts with the observed arrow of time.  Boltzmann brains ``win'', and the probability of our own observations is double-exponentially small.  (It is not exactly zero even in the first case, since a small fraction of Boltzmann brains do see a large arrow of time.)  Only in the final case do ordinary observers win, in the sense that all but a double-exponentially small fraction of observations are made by them.

Note that initial conditions are still relevant in this model.  Not only do we need to start out in the de~Sitter vacuum, but also at relatively low entropy, for example with a hot big bang.  Suppose that, by contrast, the universe started out as empty de~Sitter space with cosmological constant $\Lambda_A$.  This is the state of maximum entropy in the vacuum $A$, given by the horizon entropy, $3\pi/\Lambda_A$.  The production of any observers requires rare fluctuations that decrease the entropy.  In that case, the number of ordinary observers vanishes, $N_{\rm OO}=0$, and there are only Boltzmann brains.  

Also, the model is rather fine-tuned if it satisfies the other necessary conditions for an arrow of time, such as the smallness of the cosmological constant, $\Lambda_A$.

\subsection{Dominant History Method}
\label{sec-domhis}

Before moving on to models with more vacua, it will be useful to develop a fast method for reproducing the above analysis.   This method is a straightforward generalization of the branching-tree implementation of the causal patch cutoff~\cite{Bou06,BouYan07} reviewed in Sec.~\ref{sec-cp}.   

Recall that the branching tree was based on a somewhat arbitrary division of events into nucleation events (when the geodesic enters a new vacuum $j$) and all other events (things that happen in the causal patch in vacuum $j$).  This choice of division is useful for computing the probability distributions over observable parameters such as the cosmological constant or the amount of spatial curvature~\cite{BouHar07,BouHar10,BouLei09,BouFre10d,BouFre10e}.  In these cases the $e_i$ are determined entirely by the distribution of vacua in the landscape, and the nontrivial ingredient is the geometry of the causal patch).  The division is also useful when computing the effective size of the landscape after dynamical selection effects~\cite{BouYan07} (in which case only the $e_i$ are relevant). 

There is a deeper reason for the usefulness of this division.  Once the geodesic enters a given vacuum $i$, it can usually be assumed that a fixed sequence of events will be set in motion, independently of how this vacuum was entered.  This turns out to be a reasonable approximation in a landscape that does give rise to an arrow of time, but not in the more general setting considered here.  For example, a vacuum might produce ordinary observers if it is entered from a neighboring vacuum with much larger cosmological constant, but the same vacuum would produce only Boltzmann brains if entered from a vacuum with lower energy.  In the string landscape, the second case is extremely unlikely---this is one of the reasons why it is successful at predicting an arrow of time---but we will encounter it in some of the small landscapes considered below.

One can remedy this shortcoming, while preserving the simplicity of the branching tree formalism, by including the production of ordinary observers and/or Boltzmann brains in the branching tree.  That is, I will treat such events in the same way as the production of a new vacuum.  The production of a Boltzmann brain in vacuum $i$ can be described by a ``decay rate'' $\Gamma_{{\rm BB},i}$ and associated branching ratio $\beta_{{\rm BB},i}$.  The decay of de~Sitter vacuum $j$ to de~Sitter vacuum $k$, if it leads to the production of ordinary observers in vacuum $k$, will be described as a decay {\em sequence}, $j\to {\rm OO}\to k$.  Similarly, it will be useful to refine the notion of initial vacuum, $*$, to distinguish between initial conditions that select empty de~Sitter space, and other initial conditions, such as a hot big bang.  

In keeping with convention, however, the quantity $\Gamma_a$ will still be defined as the sum of the decay rates to other {\em vacua} (i.e., not including $\Gamma_{{\rm BB},a}$).  Therefore, the definition of branching ratios in Eq.~(\ref{eq-branch}) must be augmented as follows:
\begin{equation}
\beta_{ba}\equiv \frac{\Gamma_{ba}}{\Gamma_a+\Gamma_{{\rm BB},a}}~,
\label{eq-branch2}
\end{equation}
where the index $b$ runs over other vacua as well as Boltzmann brains ($b={\rm BB}$) in vacuum $a$.  This modification is important in the case where $\Gamma_{{\rm BB},a}> \Gamma_a$; in this case $1-\beta_{{\rm BB},a}=\Gamma_a\lll 1$ by double-exponential arithmetic, and of order $\Gamma_a^{-1}$ Boltzmann brains are produced before the vacuum decays.  In the opposite case, $\Gamma_{{\rm BB},a}< \Gamma_a$, one finds $\beta_{{\rm BB},a}=\Gamma_{{\rm BB},a}\lll 1$ from Eq.~(\ref{eq-de}).

To compute the probability of observing an arrow of time, one must compare the expected numbers of ordinary observers (who will observe such an arrow) and of Boltzmann brains (who almost certainly will not).  Therefore, we need only consider histories that include some kind of observer, even if all of those histories are quite suppressed.  The analysis is simplified in many cases where there will be a dominant history or class of histories for each type of observers.

Let us begin by applying this method to the case (a) in the previous subsection (Fig.~\ref{fig-AT}a): $\Gamma_{{\rm BB},A}<\Gamma_A$ and initial conditions are low-entropy, so that ordinary observers form.   On a timescale of order $t_\Lambda \log t_\Lambda$, all matter is diluted by the exponential de~Sitter expansion, and the causal patch becomes empty de~Sitter space.  After a time of order $\Gamma_A^{-1}$, vacuum $A$ decays to the terminal vacuum $T$, and the causal patch quickly ends in a crunch.   Thus, the dominant path is 
\begin{equation}
\mathrm{bang} \xrightarrow{1'} \mathrm{OO} ~~~[\xrightarrow{1'}  A \xrightarrow{1'} T \xrightarrow{1} \mathrm{crunch}]~,
\label{eq-casea}
\end{equation}
where $A$ denotes the empty de~Sitter phase of vacuum $A$.  Branching ratios are denoted above the arrows; the notation $1'$ means a branching ratio that is unity up to a negligible correction.  (For example, there is a small probability for decay to the terminal vacuum between the big bang and the production of the ordinary observers are produced.)  The relevant portion of this path ends with the production of ordinary observers.  (In order to present a complete causal patch, the most likely history beyond this point is included in square brackets.)  The product of branching ratios is given by the single ratio appearing between ``big bang'' and ``OO'', and is almost unity.

This result must be compared to the dominant path that leads to the production of Boltzmann brains from the same initial condition:
\begin{equation}
\mathrm{bang} \xrightarrow{1'} \mathrm{OO} \xrightarrow{1'}  A \xrightarrow{\Gamma_{{\rm BB},A}} {\rm BB} ~~~[\xrightarrow{1'}  A \xrightarrow{1'}  T \xrightarrow{1} \mathrm{crunch}]~,
\end{equation}
This and similar alternate paths all contain at least one double-exponentially small branching ratio, $\Gamma_{{\rm BB},A}$.  The product of branching ratios will be at least this small.  Comparison with Eq.~(\ref{eq-casea})  implies that Boltzmann brains are vastly outnumbered by ordinary observers.  Therefore, most observers are ordinary and see an arrow of time.

Next, consider the case shown in Fig.~\ref{fig-AT}b: the decay of $A$ is still fast ($\Gamma_A>\Gamma_{{\rm BB},A}$), but the causal patch starts as empty de~Sitter space, $A$, with large entropy.  The dominant history overall is a decay from $A$ to $T$ followed by a crunch, $A\to T\to$ crunch.  But this history has no observers of any type, so it can be ignored.  The most probable history {\em with\/} observers is the production of a Boltzmann brain in otherwise empty de~Sitter space, followed by the decay of the $A$-vacuum and a crunch:
\begin{equation}
A \xrightarrow{\Gamma_{{\rm BB},A}} \mathrm{BB} ~~~[\xrightarrow{1'} A \xrightarrow{1'} T \xrightarrow{1}  \mathrm{crunch}]~.
\label{eq-caseb}
\end{equation}
Ordinary observers can only be thermally produced, via the path
\begin{equation}
A \xrightarrow{\Gamma_{\mathrm{bang},A}} \mathrm{bang} \xrightarrow{1'} \mathrm{OO}~~~[ \xrightarrow{1'} A \xrightarrow{1'} T \xrightarrow{1} \mathrm{crunch}]~.
\end{equation}
Here, ``bang'' refers to whatever early state in standard cosmology we picked when defining the observed arrow of time.  Since the entropy of this state is exponentially smaller than that of a Boltzmann brain in empty de~Sitter space, this path is double-exponentially suppressed compared to (\ref{eq-caseb}):
\begin{equation}
\Gamma_{\mathrm{bang},A}\lll \Gamma_{{\rm BB},A}~.
\end{equation}
Therefore, most observers are Boltzmann brains and do not see an arrow of time.

Finally, consider the case of Fig.~\ref{fig-AT}c: the decay of $A$ is slow, $\Gamma_A<\Gamma_{{\rm BB},A}$.  Even though the universe begins with a low-entropy big bang leading to ordinary observers, the most likely history produces of order $\Gamma_A^{-1}$ Boltzmann brains before the vacuum $A$ finally decays:
\begin{equation}
\mathrm{bang} \xrightarrow{1'} \mathrm{OO} \xrightarrow{1'} A \xrightarrow{1'} \mathrm{BB} \xrightarrow{1'}  A \xrightarrow{1'}  \mathrm{BB} \xrightarrow{1'}  A \xrightarrow{1'} \ldots \xrightarrow{1'} \mathrm{BB}~~~ [ \xrightarrow{1'} A \xrightarrow{\Gamma_A} T \xrightarrow{1} \mathrm{crunch}]~,
\label{eq-casec}
\end{equation}
where $\ldots$ stands for $O(\Gamma_A^{-1})$ repetitions of $\to \mathrm{BB} \to A$.  The history is coarse-grained in that the exact number of Boltzmann brain creation events is irrelevant and is summed over. 

The above path is both the dominant path to the production of ordinary observers, and of Boltzmann brains.  But the expected number of Boltzmann brain production events outnumbers that for ordinary observers, by the double-exponential factor $\Gamma_A^{-1}$.\footnote{Here and in Eq.~(\ref{eq-casec}) I have implicitly assumed that $\Gamma_A^{-1}<\Gamma_{{\rm OO},A}^{-1}$.  More generally, Boltzmann brains outnumber ordinary observers by the factor $\min\{\Gamma_A^{-1},\Gamma_{{\rm OO},A}^{-1}\}$, where $\Gamma_{{\rm OO},A}^{-1}<\Gamma_{{\rm BB},A}^{-1}$ is the is the rate of production of the particularly rare Boltzmann brains that have an ordinary observer's environment (or at least his memory). Such Boltzmann brains see a long arrow of time.  The latter case applies if $\Gamma_A^{-1}>\Gamma_{{\rm OO},A}^{-1}$, i.e., the vacuum $A$ is so long lived that such events occur. Then the ``$\ldots$'' in Eq.~(\ref{eq-casec}) contains some elements of the form ``$\to \mathrm{OO} \to A$'', but with $\mathrm{BB}$ still vastly outnumbering $\mathrm{OO}$.}  The number of ordinary observers produced per event is not double-exponential.  Therefore, typical observers are Boltzmann brains, and the observation of an arrow of time has essentially zero probability.

\section{Two Landscapes with Four Vacua}
\label{sec-four}

As a preparation for our analysis of the string landscape in Sec.~\ref{sec-big}, it will be instructive to contrast two slightly larger landscapes.  In both, I will consider initial conditions that have higher entropy than any state with observers.  In the first landscape, this will lead to Boltzmann brain domination.  In the second, perhaps surprisingly, an arrow of time is predicted, capturing a key feature of the string landscape.

\begin{figure}[tbp]
\includegraphics[width=6in]{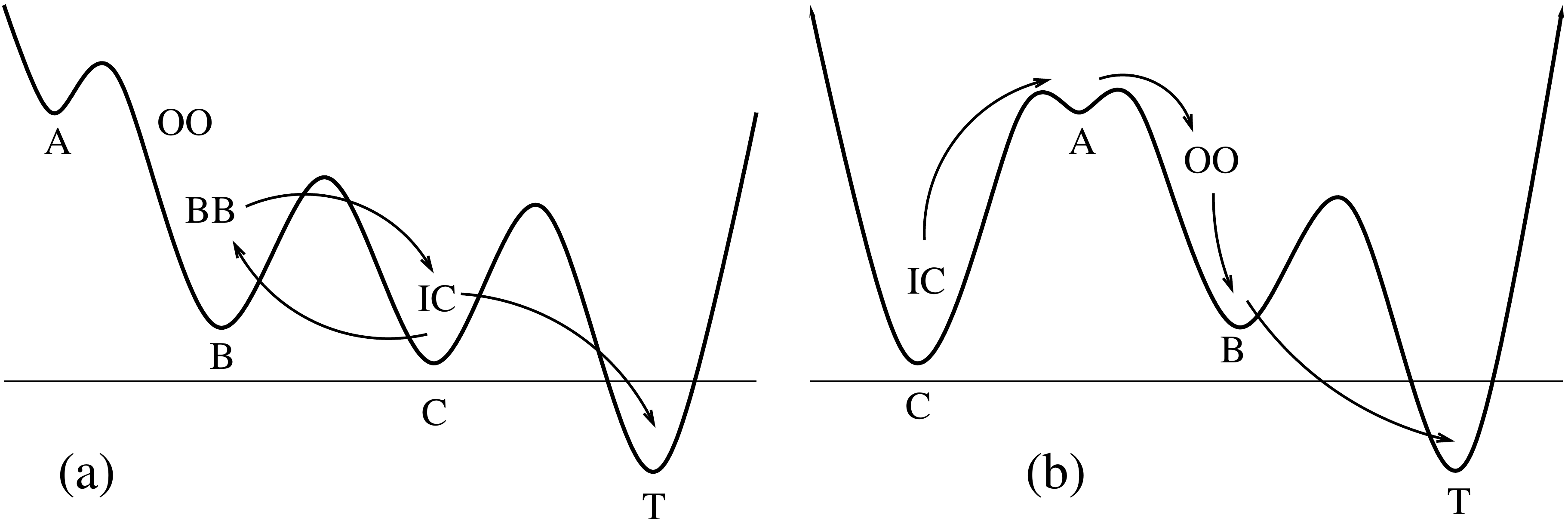}
 \caption{Two landscapes with four vacua each.  In both models, the universe starts out in vacuum $C$, and the cosmological constant and low-energy physics are the same in identically labelled vacua.   In model (a), the most likely way to produce observers is by a rare fluctuation to a Boltzmann brain state of vacuum $B$; no arrow of time is predicted.  In model (b), the only way to produce observers is by first fluctuating up to vacuum $A$; a large arrow of time is predicted even though initial conditions have arbitrarily large entropy.  The string landscape shares crucial features of this second model.}
\label{fig-ABCT} 
\end{figure}
Consider the one-dimensional landscape shown in Fig.~\ref{fig-ABCT}(a), with vacua (in descending order of $\Lambda$) $A$, $B$, $C$, and the terminal vacuum $T$.  The microphysics is assumed such that ordinary observers are produced when $A$ decays to $B$.  Neither ordinary observers nor Boltzmann brains can exist in any vacuum other than $B$.  The decay of $B$ is faster than the production of Boltzmann brains in $B$: $\Gamma_B>\Gamma_{\mathrm{BB}, B}$.  

The above assumptions imply that $\Lambda_B$ and $\Lambda_C$ are both very small: $\Lambda_C<\Lambda_B<\exp(-S_{\mathrm{BB}})$.  In Sec.~\ref{sec-big}, I will argue that this property is rare among neighboring vacua in the string landscape.

With low-entropy initial conditions in the $A$ vacuum, the dominant history would be
$A \xrightarrow{1'}  \mathrm{OO}~~ [\xrightarrow{1'} B \xrightarrow{1'} C \xrightarrow{1'} T\xrightarrow{1} \mathrm{crunch}]$, and most observers would be ordinary and would see an arrow of time. However, I will instead assume high-entropy initial conditions, in the $C$ vacuum.  Then the dominant history leading to Boltzmann brains is
\begin{equation}
C \xrightarrow{\Gamma_{BC}}  B \xrightarrow{\Gamma_{{\rm BB},B}}  \mathrm{BB} ~~~[\xrightarrow{1'}  B \xrightarrow{1'} C \xrightarrow{1'} T\xrightarrow{1} \mathrm{crunch}]~.
\label{eq-fourvac}
\end{equation}
The dominant history leading to ordinary observers is
\begin{equation}
C \xrightarrow{\Gamma_{BC}}  B \xrightarrow{\Gamma_{{\rm OO},B}}  \mathrm{OO} ~~~[\xrightarrow{1'}  B \xrightarrow{1'} C \xrightarrow{1'} T\xrightarrow{1} \mathrm{crunch}]~.
\label{eq-fourvacoo}
\end{equation}
Since $\Gamma_{{\rm OO},B}\lll \Gamma_{{\rm BB},B}$, most observers are Boltzmann brains:
\begin{equation}
e_{{\rm BB}}\ggg e_{{\rm OO}}~.
\label{eq-mod1}
\end{equation}
In this landscape, the existence of an arrow of time depends essentially on the same conditions as in the two-vacuum landscape of Sec.~\ref{sec-twovac}.  In particular, there is no arrow of time if initial conditions select a state of entropy higher than that of a state with ordinary observers, such as $C$.

Next, consider the one-dimensional landscape shown in Fig.~\ref{fig-ABCT}(b).  The cosmological constant each of the vacua $A$, $B$, $C$, and $T$ takes the same values as in the previous model.  As before, ordinary observers are produced by the decay from $A$ to $B$; $B$ is the only vacuum with any type of observers; and initial conditions will be in the vacuum $C$.

There is only one difference to the previous model: the vacua have been ordered differently, so that $B$ cannot be reached from $C$ except by first jumping up to $A$.  This means that with initial conditions in $C$, the dominant path to ordinary observers will be
\begin{equation}
C \xrightarrow{1'}  A \xrightarrow{\Gamma_{BA}} \mathrm{OO}~~~ [\xrightarrow{1'} B \xrightarrow{1'}  T \xrightarrow{1} \mathrm{crunch}]~.
\end{equation}
With the same initial conditions, the dominant path to Boltzmann brains is suppressed by an extra factor of $\Gamma_{{\rm BB},B}$:
\begin{equation}
C \xrightarrow{1'}  A \xrightarrow{\Gamma_{BA}} \mathrm{OO} \xrightarrow{1'} B \xrightarrow{\Gamma_{{\rm BB},B}} \mathrm{BB} ~~~ [\xrightarrow{1'} B \xrightarrow{1'}  T \xrightarrow{1} \mathrm{crunch}]~.
\end{equation}
Therefore, most observers are ordinary:
\begin{equation}
e_{{\rm OO}}\ggg e_{{\rm BB}}~.
\label{eq-mod2}
\end{equation}
This should be compared to Eq.~(\ref{eq-mod1}).  In both models, the initial conditions select very large entropy.  But in the latter model, the theory nevertheless predicts an arrow of time.  

This is a very simple example of a situation where the arrow of time arises dynamically despite high-entropy initial conditions.  The key feature that underlies this surprising property is the fact that there is no way to evolve from the high-entropy initial state to a state with any observers (Boltzmann or ordinary), except via the low-entropy state $A$.  In the following section, I will argue that the string landscape shares this property and thus predicts an arrow of time, no matter how large the initial entropy is.

\section{String Theory Landscape}
\label{sec-big}

Now, let us ask whether an arrow of time is predicted in the string landscape.  

\subsection{Key Properties}
\label{sec-properties}

Only the following properties of the string landscape will be used, and the conclusion below applies to any landscape that shares these properties:
\begin{itemize}

\item[(i)]{{\em No tuning.} Only a tiny fraction of vacua contain observers of any type, including Boltzmann brains.  Fine-tuned conditions that are necessary for observers, such as the smallness of the cosmological constant ($|\Lambda|< S_{\mathrm{BB}}^{-1}$), or nontrivial low-energy physics, arise accidentally in some vacua, because the total number of vacua is very large.}

\item[(ii)]{{\em Large step size.} Vacuum transitions generically change the cosmological constant by an amount $|\Delta \Lambda| \gg S_{\mathrm{BB}}^{-1}$.}

\item[(iii)]{{\em Not too large.} The effective number of vacua is less than $\exp(S_{\mathrm{BB}})$.}

\item[(iv)]{{\em Not effectively one-dimensional.} For any two de~Sitter vacua $a$, $b$, with $\Lambda_b< S_{\mathrm{BB}}^{-1}$, there exists a semiclassical decay path from $a$ to $b$ that does not pass through any vacuum $i$ with $\Lambda_i< S_{\mathrm{BB}}^{-1}$.}

\end{itemize}

Here $S_{{\rm BB}}$ is the minimum coarse-grained entropy of a Boltzmann brain.  In general this number need not be the same in all vacua that can produce Boltzmann brains; however, its precise value is irrelevant as long as it is exponentially large.  For definiteness, we could take $S_{{\rm BB}}\sim 10^{25}$~\cite{Ban07}.  Because of entropy bounds in de~Sitter space~\cite{Bou00b}, the production rate of Boltzmann brains in all vacua satisfies~\cite{BouFre06b,BouFre08b}
\begin{equation}
\Gamma_{{\rm BB},a}<\exp(-S_{{\rm BB}})~.
\label{eq-gbbsbb}
\end{equation} 
Another way of thinking about this restriction is that a Boltzmann brain requires a minimum free energy (in the above guess, of order $10^{25}$), in order to observe and process information.  Its production will be correspondingly suppressed.  

A key point is that the overwhelming majority of Boltzmann brains will at most observe the conversion of their own free energy into entropy, since the spontaneous reduction of entropy by more than the minimum amount needed to produce a Boltzmann brain is exponentially suppressed.  They will not observe a large arrow of time, i.e., the production of a large amount $\Delta S\sim 10^{103}$ of entropy, as we do; see Eq.~(\ref{eq-at}).

Of course, I will also assume that the landscape contains ordinary observers in some de~Sitter vacua; otherwise the theory is trivially ruled out.  More precisely, I assume that the expected number of ordinary observers that are produced by in some vacua by the decay of some other vacua is not double-exponentially small.  I will now discuss each of the above landscape properties in turn.

The first property---no tuning---is desirable from an aesthetic viewpoint, since the alternative would be to posit some form of intelligent design of the effective laws of physics.  However, this is not why it is listed here.  Rather, as we shall see below, the {\em absence\/} of tuning actually {\em helps\/} explain the observed arrow of time.---When combined with simple observations, the no-tuning property implies a lower bound on the number of vacua in the landscape.  The fraction of vacua that have cosmological constant comparable to the observed magnitude is of order $10^{-123}$; and the fraction of those vacua that contain observers will be some other number $\epsilon<1$.  One expects that $\epsilon$ is exponentially small, since the complexity observed in our own vacuum does not appear to be robust against small variations of several parameters~\cite{BarTip,HalNom07,BouHal09}, such as the strong and electroweak coupling strengths; the masses of the electron, proton, and neutron; and the amplitude of primordial density perturbations.  For the landscape to be compatible with a vacuum like ours, without tuning, the total number must be at least of order $\epsilon^{-1} 10^{123}$.   The string landscape appears to satisfy this condition~\cite{BP,KKLT,DenDou04b}.

The second property---large step size---could be considered a consequence of the first, since a landscape with small step size~\cite{Abb85,BT1,BT2} would seem to require a small input parameter $\Delta \Lambda\ll 1$.  In any case, it is worth spelling this requirement out, because of its crucial rule in the arguments below.  Note that the string landscape is expected to satisfy this property with much room to spare, $\Delta \Lambda\lesssim 1$.  A very similar condition, $|\Delta \Lambda|\gg G_N \rho_{\rm BBN}\sim 10^{-88}$, where $\rho_{\rm BBN}$ is the energy density at the time of big bang nucleosynthesis, is necessary for solving the cosmological constant problem without predicting an empty universe~\cite{BP,Pol06,TASI07}.

The third property, that the landscape contains fewer than $\exp(S_{\mathrm{BB}})$ vacua, is also expected to hold in the string landscape, whose number of vacua is controlled, roughly, by the exponential of the number of topological cycles of a six-dimensional compact manifold~\cite{BP,KKLT,DenDou04b}, ${\cal N} \sim 10^{O(100-1000)}$.  Larger numbers may arise from F-theory compactifications~\cite{DenDou06}, but they are still much smaller than $\exp(S_{{\rm BB}})$.  By ``effective'', I mean the number of vacua that are sufficiently likely to be dynamically produced, in the sense of Ref.~\cite{BouYan07}.  This presumably excludes the (infinite) number of stable AdS vacua of string theory, whose large fluxes will be difficult to produce cosmologically.  

The fourth property states that there is more than one way to get from one vacuum to another by a sequence of Coleman-de Luccia tunneling events. Imagine a large two-dimensional landscape of vacua; generically, any pair of de~Sitter vacua is connected by many different decay chains.   The potential landscape of string theory is believed to have hundreds of dimensions~\cite{BP} and is thus expected to satisfy this property.  The fourth property would {\em not\/} hold in a one-dimensional landscape such as the Abbott~\cite{Abb85} or Brown-Teitelboim~\cite{BT1,BT2} models, or in a landscape that had a ``bottleneck''---a vacuum with small $\Lambda$ that connects two otherwise disconnected portions of the landscape.  (It is not necessary to assume that the whole landscape is connected, i.e., that there exists a decay path connecting any two vacua.  This is automatic, since we take the effective landscape to be the portion accessible from the given initial conditions, and we demand that the above properties hold for this portion alone.) 

\subsection{Arrow of Time}
\label{sec-proof}

Using the above four properties,\footnote{The assumptions, claim, and proof presented in this section all differ slightly, but not substantially, from those presented in Ref.~\cite{BouFre08b,DesGut08b}.} I will now argue that {\em an arrow of time is predicted if and only if all de~Sitter vacua decay faster than they produce Boltzmann brains}, i.e., if the condition
\begin{equation}
\Gamma_{{\rm BB},a} < \Gamma_a
\label{eq-condition}
\end{equation}
holds for all de~Sitter vacua $a$.  {\em The initial entropy is irrelevant.}

Before presenting a proof, it is instructive to give an intuitive sketch of the argument, based on our experience with the models of Sec.~\ref{sec-four}.  To predict an arrow of time, the string landscape should be, in a sense, less like the first and more like the second landscape studied there.  This is indeed ensured by the four properties listed.  

Eqs.~(\ref{eq-gbbsbb}) and (\ref{eq-condition}) guarantee that Boltzmann brains are suppressed at least by a branching ratio of order $\exp(-S_{{\rm BB}})$.  Vacua with small enough cosmological constant to contain ordinary observers satisfy $\Lambda<S_{{\rm BB}}^{-1}$; otherwise, the maximum entropy is too small even for a Boltzmann brain to fit.  By the large-step-size condition (ii), such vacua can only be reached by down-tunneling from a vacuum of much larger cosmological constant.  (Up-tunneling from vacua with negative $\Lambda$ is semi-classically forbidden.)  This excludes the type of problem that arose in the first model of Sec.~\ref{sec-four}, where the vacuum with ordinary observers could only be reached by up-tunneling from a de~Sitter vacuum with even smaller $\Lambda$, producing a state of maximum entropy from which observers could only arise by Boltzmann-suppressed fluctuations.  The third and fourth property ensure that the paths that do lead to ordinary observers are less suppressed than $\exp(-S_{{\rm BB}})$, so that ordinary observers dominate.

An exception would occur if the initial conditions selected a high-entropy state in a de~Sitter vacuum that can produce Boltzmann brains, such as case (b) in Fig.~\ref{fig-AT}.  This is (statistically) excluded by the first condition: vacua capable of producing Boltzmann brains are exponentially rare, so whatever theory selects the initial condition would have to be tuned to pick such a vacuum.

If, on the other hand, Eq.~(\ref{eq-condition}) did not hold in some vacuum, then by the same analysis as in case (b) Sec.~\ref{sec-two}, a double-exponentially large number of Boltzmann brains would be produced in that vacuum; see Eq.~(\ref{eq-casec}).  Since at most an exponentially large number of ordinary observers is produced in any vacuum by property (iii), an arrow of time would not be predicted.  Thus Eq.~(\ref{eq-condition}) is necessary for an arrow of time.

\paragraph{Proof of ``if''} Let us assume that $\Gamma_{{\rm BB},a} < \Gamma_a$ for all de~Sitter vacua $a$ whose microphysics allows, in principle, for Boltzmann brains.   By Eq.~(\ref{eq-gbbsbb}), $\Gamma_{{\rm BB},a}$ is double-exponentially small.  The absence of tuning, property (i), and double-exponential arithmetic, Eq.~(\ref{eq-de}), imply that any path to Boltzmann brains is suppressed at least by $\exp(-S_{{\rm BB}})$, the branching ratio to Boltzmann brains in the vacuum $a$.  There may be many such paths with comparable weight, but not double-exponentially many, by condition (iii).  So this is the minimum suppression of Boltzmann brains, to the accuracy demanded by double-exponential arithmetic:
\begin{equation}
e_{\rm BB}<\exp(-S_{{\rm BB}})~.
\end{equation}

Boltzmann brains would nevertheless dominate over ordinary observers if the initial conditions select an empty de~Sitter vacuum, $*$, that is capable of producing Boltzmann brains~\cite{BouFre08b,DesGut08b}.  (This is shown in the Appendix.)  However, this possibility conflicts with condition (i).  Vacua containing Boltzmann brains are extremely rare.  One could imagine that the theory of initial conditions favors some of the more primitive necessary conditions for Boltzmann brains, such as a small cosmological constant.  However, many additional conditions are necessary for the low energy physics to be compatible with the level of complexity required for the operation of a Boltzmann brain.  The theory of initial conditions would have to be tuned to select for a vacuum with observers, in violation of (i).\footnote{In this paper, it is assumed that initial conditions have support in one vacuum only.  The general case is left to future work.}

Given that the initial vacuum does not support Boltzmann brains, it will now be shown that the expected number of ordinary observers is greater than that of Boltzmann brains.  There are two cases, depending on the initial value of the cosmological constant.  If $\Lambda_*<S_{{\rm BB}}^{-1}$, then the large-step-size property (ii) implies\footnote{Property (ii) is a statistical statement about the distribution of cosmological constant among vacua.  It implies that among all vacua that satisfy $\Lambda<S_{{\rm BB}}^{-1}$, a fraction no greater than $S_{{\rm BB}}^{-1}/|\Delta \Lambda|$ decay directly to vacua that satisfy the same property.  Thus, the initial conditions would have to be finetuned for this to be the case for the vacuum $*$, in violation of property (i).} that any path---whether it leads to ordinary observers or to Boltzmann brains---must begin with up-tunneling from $*$ to some vacuum $i_1$ with cosmological constant $\Lambda_{i_1}\gg S_{{\rm BB}}^{-1}$, so that 
\begin{equation}
\exp(S_{i_1})\lll \exp(S_{\rm BB})
\label{eq-abcl}
\end{equation}
where $S_{i_1}=3\pi/\Lambda_{i_1}$ is the de~Sitter entropy of the vacuum $i_1$.
Detailed balance relates the up-tunneling rate, $\Gamma_{i_1*}$, to the decay rate back down to the $*$ vacuum, $\Gamma_{*i_1}$: 
\begin{equation}
\Gamma_{i_1*}\exp(S_*)=\Gamma_{*i_1}\exp(S_{i_1})~.
\label{eq-abcn}
\end{equation}
Moreover, the decay time of vacuum $i_1$ cannot be larger than the recurrence time~\cite{DysKle02} or faster than the Planck time: 
\begin{equation}
1>\Gamma_{*i_1}>\exp(-S_{i_1})~.  
\end{equation}
Substituting on the right hand side of Eq.~(\ref{eq-abcn}), this implies the double inequality
\begin{equation}
\exp(S_{i_1})>\Gamma_{i_1*}\exp(S_*)> 1~.
\label{eq-cba}
\end{equation}
Eq.~(\ref{eq-abcl}) now implies that to accuracy better than $\exp(S_{{\rm BB}})$, this first uptunneling suppresses the paths by the same amount, whether they lead to Boltzmann brains or to ordinary observers:
\begin{equation}
\exp(-S_{{\rm BB}})\lll \frac{\beta_{i_{1,{\rm OO}}*}}{\beta_{i_{1,{\rm BB}}*}} = \frac{\Gamma_{i_{1,{\rm OO}}*}}{\Gamma_{i_{1,{\rm BB}}*}}\lll \exp(S_{{\rm BB}})~.
\end{equation}
(The distinction between $i_{1,{\rm BB}}$ and $i_{1,{\rm OO}}$ is important since the dominant paths leading to Boltzmann brains and those leading to ordinary observers may prefer different $i_1$.)  

This accuracy is sufficient, since it will be shown that ordinary observers are favored over Boltzmann brains by a factor greater than $\exp(S_{{\rm BB}})$.  Thus, we can forget about this first case altogether: effectively, the cost of first up-tunneling, which leads to a ``new initial vacuum'' with much larger cosmological constant, cancels between paths to ordinary observers and paths to Boltzmann brains.  The rest of the argument proceeds as in the second case, with the vacuum $i_{1,{\rm OO}}$ playing the role of $*$ below.

The second case is that of ``large'' initial cosmological constant, in the sense that $\Lambda_*>S_{{\rm BB}}^{-1}$.   By property (iv), there will exist a path to a vacuum with ordinary observers consisting entirely of vacua $i_2, i_3, \ldots$ with $\Lambda_{i_n}>S_{{\rm BB}}^{-1}$ for all $n\geq 2$.   The decay channels of these vacua have rates that vary between $1$ and $\exp(-S_{i_n})$; hence, all branching ratios appearing in the path are larger than $\exp(-S_{{\rm BB}})$.  By property (iii), the paths are not double-exponentially long, so it follows that the overall weight of paths leading to ordinary observers exceeds $\exp(-S_{{\rm BB}})$.  

Combining these results, one may bound the weights, or unnormalized probabilities, as follows:
\begin{eqnarray}
e_{{\rm OO}} & > & X_{{\rm OO}} \exp(-S_{{\rm BB}})~,\\
e_{{\rm BB}} & < & X_{{\rm BB}} \exp(-S_{{\rm BB}})~,
\end{eqnarray}
where $X_{{\rm OO}}$ differs from $X_{{\rm BB}}$ by less than a factor of $\exp(\pm S_{{\rm BB}})$.  Let us recall the trivial assumption that there exist vacua in which decay of a parent vacuum produces a number of ordinary observers that is not double-exponentially small.  Using Eq.~(\ref{eq-de}) and (\ref{eq-branch}), this implies that there are far more ordinary observers than Boltzmann brains.  Thus, an arrow of time is predicted.

\paragraph{Proof of ``only if''} Suppose that a vacuum $A$ exists with $\Gamma_{{\rm BB},A} > \Gamma_A$.  Then the branching ratio to Boltzmann brains is nearly unity once this vacuum is reached.  Of order $\Gamma_{{\rm BB},A}/\Gamma_A\approx \Gamma_A^{-1}$ Boltzmann brains will be produced in this path:
\begin{equation}
\ldots \xrightarrow{1'} A \xrightarrow{1'} \mathrm{BB} \xrightarrow{1'}  A \xrightarrow{1'}  \mathrm{BB} \xrightarrow{1'}  A \xrightarrow{1'} \ldots \xrightarrow{1'} \mathrm{BB}~~~ [ \xrightarrow{1'} A \xrightarrow{\Gamma_A} T \xrightarrow{1} \mathrm{crunch}]~,
\end{equation}
This number is double-exponentially large.  

On the other hand, the expected number of ordinary observers is at most exponentially large.  This follows from properties (i) and (iii), which ensure that the smallest positive $\Lambda$ in the landscape is exponentially but not double-exponentially small.  The entropy bound in the corresponding de~Sitter space is $3\pi/\Lambda$, and the number of observers cannot be larger than the entropy.  Therefore, their number cannot be double-exponentially large in any vacuum.  

As before, the other branching ratios along any dominant path are large compared to $\exp(-S_{{\rm BB}})$, up to a possible suppression from up-tunneling out of the $*$ vacuum that is common to both Boltzmann brains and ordinary observers.  Thus, the condition
\begin{equation}
\Gamma_{{\rm BB},a} < \Gamma_a~,~~ \mathrm{for~all}~a~,
\end{equation}
is necessary for an arrow  of time.

\acknowledgments  I would like to thank A.~Brown, B.~Freivogel, M.~Kleban, S.~Leichenauer, I.~Yang and C.~Zukowski for valuable discussions and comments.  I am indebted to D.~Page for discussions of the implications of the present results for the viability of the Hartle-Hawking proposal. This work was supported by the Berkeley Center for Theoretical Physics, by the National Science Foundation (award numbers 0855653 and 0756174), by fqxi grant RFP3-1004, and by the US Department of Energy under Contract DE-AC02-05CH11231.

\appendix

\section{Initial vacuum with observers}

Here I will show that a large arrow of time is {\em not\/} predicted in the landscape of Sec.~\ref{sec-big}, if the initial vacuum, $*$, has a nonzero rate
\begin{equation}
\Gamma_{{\rm BB},*}>0
\label{eq-jj}
\end{equation}
for producing Boltzmann brains.  In the proof of Sec.~\ref{sec-proof}, it is argued that this case will not arise in the absence of tuning.  The purpose of this appendix is merely to explain why this case did need to be excluded by some argument.  (A closely related analysis appears in Refs.~\cite{BouFre08b,DesGut08b}.)

By Eq.~(\ref{eq-gbbsbb}), the above assumption would require, in particular, that the initial vacuum has very small cosmological constant,
\begin{equation}
\Lambda_*<\exp(-S_{{\rm BB}})~,
\label{eq-lkj}
\end{equation}
so all paths to other de~Sitter vacua begin by an up-tunneling to a vacuum $i_1$ with larger cosmological constant.  

By Eq.~(\ref{eq-condition}) and double-exponential arithmetic, the unnormalized probability for Boltzmann brains is at least
\begin{equation}
e_{{\rm BB}}\geq \Gamma_{{\rm BB},*}~,
\end{equation}
 from the path
\begin{equation}
* \xrightarrow{\Gamma_{{\rm BB},*}} {\rm BB}~~~ [ \xrightarrow{1'} T \xrightarrow{1} \mathrm{crunch}] ~.
\end{equation}
($e_{{\rm BB}}$ could be larger than $\Gamma_{{\rm BB},*}$, if $\Gamma_{{\rm BB},*}$ is so small that a path involving tunneling to a different vacuum with Boltzmann brains dominates.)

The dominant path to ordinary observers is suppressed by 
\begin{equation}
e_{{\rm OO}}=\Gamma_{{\rm OO},*}\lll e_{{\rm BB}}~,
\end{equation}
from the dominant path
\begin{equation}
* \xrightarrow{\Gamma_{{\rm OO},*}} {\rm OO}~~~ [ \xrightarrow{1'} T \xrightarrow{1} \mathrm{crunch}] ~,
\end{equation}
{\em if\/} such observers exist in the initial vacuum.  If they do not, then the dominant path to ordinary observers involves up-tunneling to some intermediate vacuum $i_{1,{\rm OO}}$.  By Eq.~(\ref{eq-abcl}) and the second inequality of Eq.~(\ref{eq-cba}), one finds for this case that
\begin{equation}
e_{{\rm OO}}< \Gamma_{i_{1,{\rm OO}}*}\lll e^{S_{{\rm BB}}}/e^{S_*}\approx \exp(-S_*)\lll \Gamma_{{\rm BB},*} < e_{{\rm BB}}
\end{equation}
Since the number of ordinary observers produced $N_{{\rm OO}}$ produced by this path is at most exponentially large by property (iii), Boltzmann brains dominate and no arrow of time is predicted.

\bibliographystyle{utcaps}
\bibliography{all}
\end{document}